\documentclass[aps,prl,twocolumn,groupedaddress,nofootinbib]{revtex4-1}

\usepackage{graphicx}
\usepackage{mathrsfs}
\newcommand{\dmt}{\delta{m}}
\newcommand{\scri}{\mathscr{I}}
\usepackage{color}

\begin{document}


\title{Exploring the small mass ratio binary black hole merger\\
via Zeno's dichotomy approach}
\thanks{The title refers to our approach of halving and halving
the mass ratio while adding internal grid refinement levels to the 
well studied $q=1/15$ case, inspired by the first of Zeno's paradoxes,
as reported by Aristotle~\cite{2000phys.book.Aristotle}.\\
J.L.Borges~\cite{Borges}, in {\it Death and the Compass}, writes
''I know of a Greek labyrinth which is a single straight line. 
Along this line so many philosophers have lost themselves...''}


\author{Carlos O. Lousto}
\author{James Healy}
\affiliation{Center for Computational Relativity and Gravitation,
School of Mathematical Sciences,
Rochester Institute of Technology, 85 Lomb Memorial Drive, Rochester,
New York 14623}


\date{\today}

\begin{abstract}
We perform a sequence of binary black hole simulations with increasingly
small mass ratios, reaching to a 128:1 binary that displays 13 orbits before
merger. Based on a detailed convergence study of the $q=m_1/m_2=1/15$ nonspinning
case, we apply additional mesh refinements levels around the smaller hole 
horizon to reach successively the $q=1/32$, $q=1/64$, and $q=1/128$ 
cases.
Roughly a linear 
computational resources  
scaling with $1/q$ is observed on 8-nodes simulations. 
We compute the remnant properties of the merger: final mass, spin, and recoil velocity, finding precise consistency between horizon and radiation measures.
We also compute the gravitational waveforms: its peak frequency, amplitude,
and luminosity. We compare those values with predictions of the
corresponding phenomenological formulas, reproducing the particle
limit within 2\%, and we then use the new results to improve their fitting 
coefficients.
\end{abstract}


\maketitle

\section{Introduction\label{sec:introduction}}

While ground based gravitational wave detectors like LIGO 
\cite{LIGOScientific:2019fpa} are particularly
sensitive to comparable (stellar) mass binaries,
third generation (3G) ground detectors~\cite{Purrer:2019jcp} and
 space detectors,
like LISA, will also be sensitive to the observation of very 
unequal mass binary black holes \cite{Gair:2017ynp}. 
These will allow the search and study of intermediate mass black holes,
either as the large hole in a merger with a stellar mass black hole
(a source for 3G detectors) or
as the smaller hole in a merger with a supermassive black hole
(a source for LISA).
The evolution of these small mass ratio
binaries has been approached via perturbation theory and the computation
of the gravitational self-force exerted by the field of the small 
black hole on itself \cite{Barack:2018yvs}.
The resolution of the binary black hole problem in its full nonlinearity
have been only possible after the 2005 breakthroughs in numerical relativity
\cite{Pretorius:2005gq,Campanelli:2005dd,Baker:2005vv},
and a first proof of principle have been performed in 
\cite{Lousto:2010ut} for the 100:1 mass ratio case,
following studies of the 10:1 and 15:1  \cite{Lousto:2010qx} ones.
In the case of \cite{Lousto:2010ut} 
the evolution covered two orbits 
before merger, and while this proved that evolutions are possible,
practical application of these gravitational waveforms
requires longer evolutions. Other approaches to the small mass ratio
regime have recently been followed \cite{Rifat:2019ltp,vandeMeent:2020xgc}.
Here we report on a new set of evolutions that are based on the 
numerical techniques
refined for the longterm evolution of a spinning precessing binary with 
mass ratio $q=m_1/m_2=1/15$~\cite{Lousto:2018dgd}.
We study here the case of a
nonspinning $q=1/15$ binary in a convergence sequence
to assess numerical and systematical errors. We then
add a sequence of $q=1/32$, $q=1/64$, and $q=1/128$, nonspinning 
binaries evolutions for about a dozen orbits before merger. 

\section{Simulations' Results\label{sec:results}}

For the $q=1/15$ case,
we performed three globally increasing resolution simulations labeled by its
number of points per total mass $m=m_1+m_2$,  n084, n100, n120,
at the waveform extraction zone, about $100-150m$ away from the binary.
All extracted waveforms are
then extrapolated to infinity ($\scri^+$) using Ref.~\cite{Nakano:2015pta} 
formulas.  The medium resolution 
''n100" simulation has 40 grid points across a radius of $0.08m$
for a finest resolution of $m/512$ at the innermost refinement level around the smaller hole. 
The whole grid consists of 12 refinement levels with an outer boundary
at $400m$ and in the wavezone, the resolution is $m/1.0$. 
The n084 (n120) simulation has globally decreased (increased) resolution 
of $m/430.1$ ($m/614.4$) at the finest level. 
The simulations start at a coordinate separation $D=8.5m$, or about
a simple proper horizon distance (along the coordinate line joining the holes),
SPD=10.1$m$. The inspiral evolution follows for about 10 orbits 
(about $t=1350m$) before
merger and forms a final black hole with the characteristics summarized
in Table~\ref{tab:q15}. The n100 simulation proceeded at a speed of
$2.2m$/hr on 8 of TACC's (\url{https://www.tacc.utexas.edu})
stampede2 nodes, costing approximately 5,340 node hours.
To compute black holes masses and spins we use the isolated horizons
technique~\cite{Dreyer:2002mx}, that produces very accurate results compared to
radiative computations~\cite{Healy:2020iuc}.

\begin{figure}
\includegraphics[angle=0,width=1.0\columnwidth]{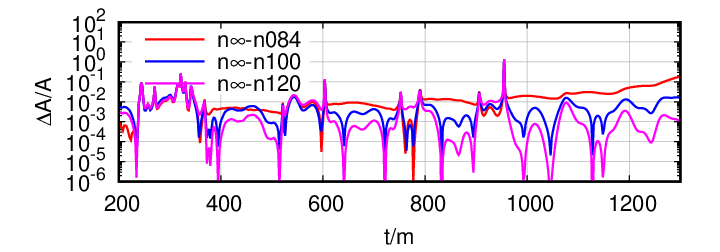}\\
\includegraphics[angle=0,width=1.0\columnwidth]{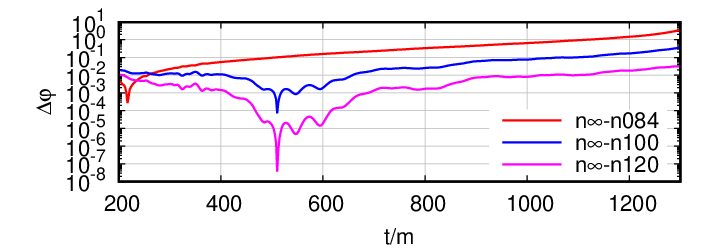}
  \caption{ Difference between each resolution of the $q=1/15$
   strain waveform with the calculated infinite resolution waveform 
   for the amplitude and phase of the (2,2) mode.
  \label{fig:q15_wf} }
\end{figure}

Table~\ref{tab:q15} displays the convergence rates for all the radiative
quantities derived from the gravitational waveform.
The detail of the convergence of the waveform phase and amplitude with global
numerical resolution is shown in Fig.~\ref{fig:q15_wf} displaying an 8th 
order convergence rate (alignment not enforced).

\begin{table*}
  \caption{
The energy radiated, $E_{\rm rad}/m$, angular momentum radiated $J_{\rm rad}/m^2$, recoil velocity $v_m$,
and the peak luminosity ${\cal L}_{\rm peak}$, waveform frequency $\omega_{22}^{\rm peak}$ at the maximum amplitude $h_{\rm peak}$, for each resolution of the $q=1/15$ simulations, starting at SPD=10$m$. All quantities are calculated from the gravitational waveforms.  Extrapolation to
infinite resolution and order of convergence is derived.
\label{tab:q15}}
\begin{ruledtabular}
\begin{tabular}{lcccccc}
resolution & $E_{\rm rad}/m$ & $J_{\rm rad}/m^2$ & $v_m$[km/s] & ${\cal L}_{\rm peak}$[ergs/s] & $m\omega_{22}^{\rm peak}$ & $(r/m)h_{\rm peak}$\\
\hline
n084         & 0.002366 & -0.029385 & 31.45 & 1.585e+55 & 0.2906 & 0.08471\\
n100         & 0.002418 & -0.029945 & 33.54 & 1.649e+55 & 0.2863 & 0.08485\\
n120         & 0.002436 & -0.030097 & 34.24 & 1.665e+55 & 0.2882 & 0.08489\\
n$\to\infty$ & 0.002444 & -0.030148 & 34.56 & 1.670e+55 & 0.2897 & 0.08489\\
order        &     6.19 &      7.58 &  6.41 &      8.11 &   4.71 &    8.83\\
\end{tabular}
\end{ruledtabular}
\end{table*}

The high convergence of these results allow us to use the medium of
the resolutions, n100, as the reference grid to perform a set of smaller 
$q$ simulations, each time halving the mass ratio and adding a new
refinement level around the smaller hole (12, 13, 14, 15 levels for 
$q= 1/15, 1/32, 1/64, 1/128$, respectively). This basic grid configuration
ensures exactly the same accuracy around the larger hole, the radiation
and boundary zones, and in between the holes, while doubling the resolution
around the smaller hole in such a way that the number of points per horizon
remains the same and guarantees its accuracy. We monitor this assumption
by verifying the conservation of the small horizon mass (and spin) to 
within the required accuracy of about one part in $10^4$.
In order to maintain the accuracy at this base
resolution and account for the longer merger time scale we reduce the 
initial distance (and hence the evolution time) as shown in 
Table~\ref{tab:qqq}. This table displays the final black hole remnant 
and peak waveform properties, and the consistency between the horizon
measures of the final mass and spin with the energy and angular momentum
carried out by the gravitational waveforms.

\begin{table*}
  \caption{
Final properties for the sequence of the $q=1/15, 1/32, 1/64, 1/128$ simulations includes 
the final black hole mass $M_{\rm rem}/m$ and spin $\alpha_{\rm rem}$, the recoil velocity $v_m$,
and the peak luminosity ${\cal L}_{\rm peak}$ and waveform frequency $\omega_{22}^{\rm peak}$ 
at the maximum amplitude $h_{\rm peak}$.
Also given are the initial simple proper distance, SPD, number of orbits to merger 
$N$, and a consistency check of the differences between the final mass and 
spin, $\Delta M_{\rm rem}/m$, $\Delta \alpha_{\rm rem}$, calculated from the horizon and from 
the radiated energy and angular momentum.
\label{tab:qqq}}
\begin{ruledtabular}
\begin{tabular}{lcccccccccc}
$q$ & $M_{\rm rem}/m$ & $\Delta M_{\rm rem}/m$ & $\alpha_{\rm rem}$ & $\Delta\alpha_{\rm rem}$ & $v_m$[km/s] & ${\cal L}_{\rm peak} [ergs/s]$ & $m\omega_{22}^{\rm peak}$ & $(r/m)h_{\rm peak}$ & SPD$/m$ & N\\
\hline
1/15   & 0.9949 & $9\times10^{-5}$ & 0.1891 & $2.3\times10^{-4}$ & 34.24 & 1.665e+55 & 0.2882 & 0.0849 & 10.13 & 10.01\\
1/32   & 0.9979 & $3\times10^{-5}$ & 0.1006 & $2.5\times10^{-3}$ &  9.14 & 4.260e+54 & 0.2820 & 0.0424 &  9.51 & 13.02\\
1/64   & 0.9990 & $5\times10^{-7}$ & 0.0520 & $2.8\times10^{-4}$ &  2.34 & 1.113e+54 & 0.2812 & 0.0220 &  8.22 &  9.98\\
1/128  & 0.9996 & $4\times10^{-5}$ & 0.0239 & $2.7\times10^{-3}$ &  0.96 & 3.313e+53 & 0.2746 & 0.0116 &  8.19 & 12.90\\

\end{tabular}
\end{ruledtabular}
\end{table*}

\begin{figure*}
\includegraphics[angle=0,width=2.0\columnwidth]{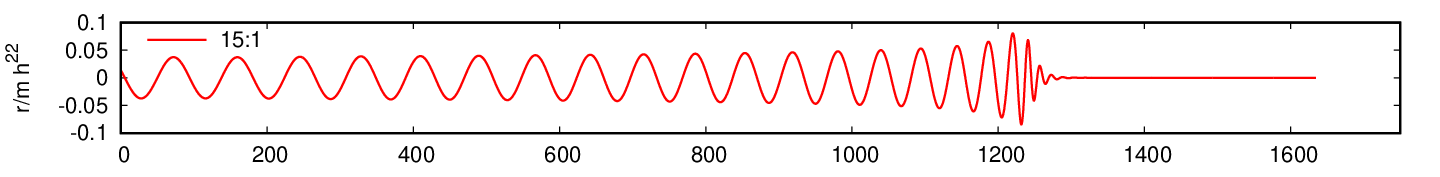}\\
\includegraphics[angle=0,width=2.0\columnwidth]{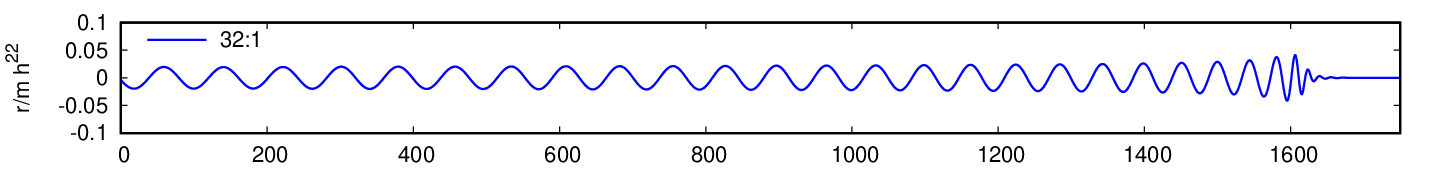}\\
\includegraphics[angle=0,width=2.0\columnwidth]{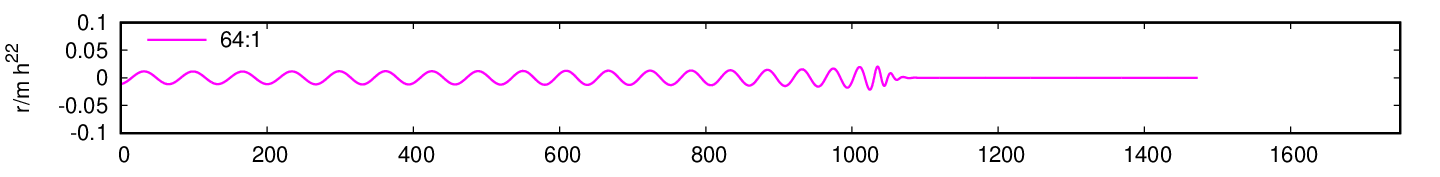}\\
\includegraphics[angle=0,width=2.0\columnwidth]{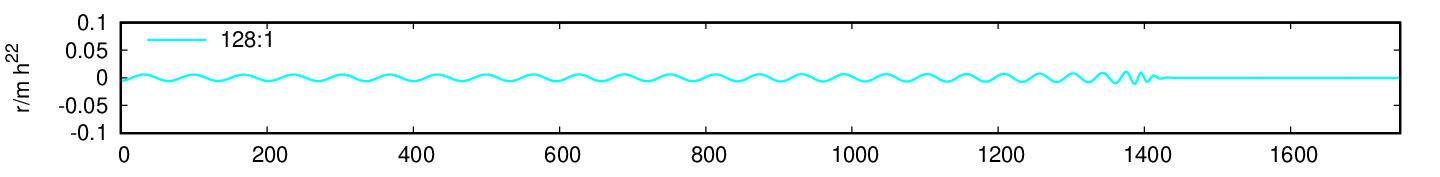}
  \caption{(2,2) modes (real part) of the strain waveforms versus time $(t/m)$, for the $q=1/15, 1/32, 1/64, 1/128$ simulations.
  \label{fig:Allq}}
\end{figure*}

\begin{figure}
\includegraphics[angle=0,width=0.95\columnwidth]{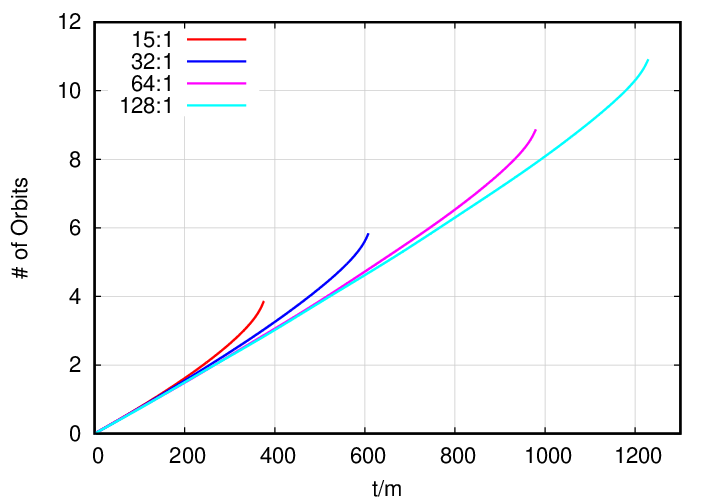}
  \caption{Comparative number of orbits and time to merger,
from a fiducial orbital frequency  $m\Omega_i=0.0465$
for the $q=1/15, 1/32, 1/64, 1/128$ simulations. 
  \label{fig:Nvst}}
\end{figure}

Figure~\ref{fig:Allq} makes a comparative display of the four $q$ 
waveforms, (2,2)-modes of the strain, in the 
same scale to show the differences in merger amplitude and evolution
time. While Fig.~\ref{fig:Nvst} displays the comparative merger time from
a fiducial initial orbital frequency $m\Omega_i=0.0465$
(corresponding roughly to coordinate
separation $D=7m$ and simple proper distance $SPD=8.5m$)
to merger for the mass ratios $q=1/15, 1/32, 1/64, 1/128$ simulations.
We observe a time to merger $t_m\sim (83.2\pm8.5)m\,\eta^{-0.56\pm0.03}$
dependence for small mass ratios, and interpret it as a composed power of the
leading rates from the post-Newtonian regime \cite{Kidder:1993zz}: 
$q^{-1}$ from the inspiral decay and $q^0$ from the plunge.

The simulations of $q=1/32, 1/64, 1/128$ use an 8th order stencil in space
\cite{Lousto:2007rj}
and 4th order in time (with $dt=dx/4$) and have all been performed in
TACC's Frontera cluster on 8 nodes (448 cores) at speeds of 1.1, 0.6, and 0.32
$m$ per hour totaling 13,807, 17,713, and 41,250 node hours respectively.
Thus showing a notable approximately linear scaling with
the mass ratio.

\section{Comparisons with Predictions\label{sec:comparisons}}

\begin{figure*}
\includegraphics[angle=0,width=0.98\columnwidth]{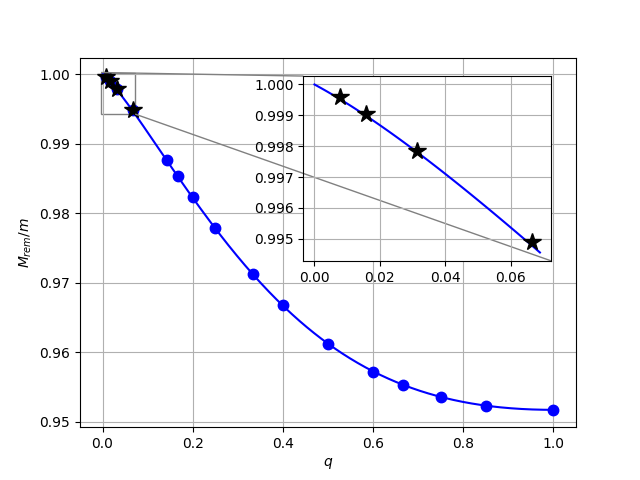}
\includegraphics[angle=0,width=0.98\columnwidth]{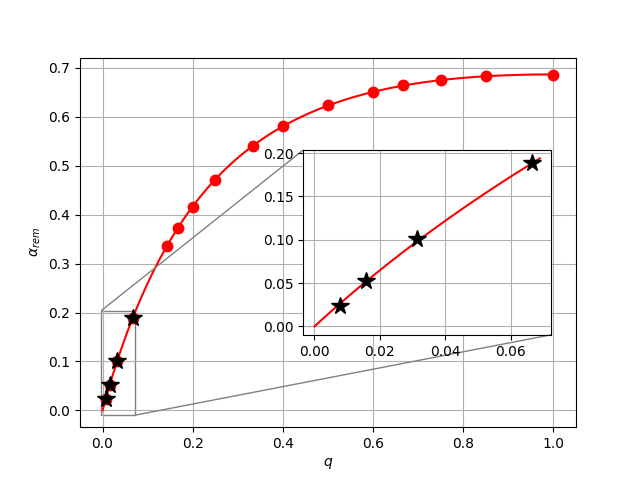}
\includegraphics[angle=0,width=0.98\columnwidth]{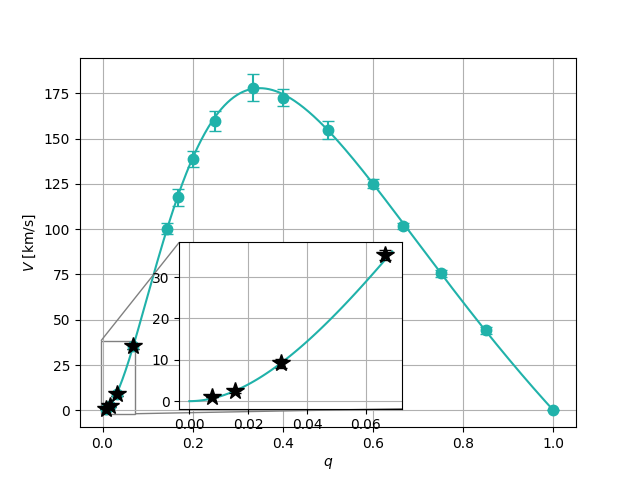}
\includegraphics[angle=0,width=0.98\columnwidth]{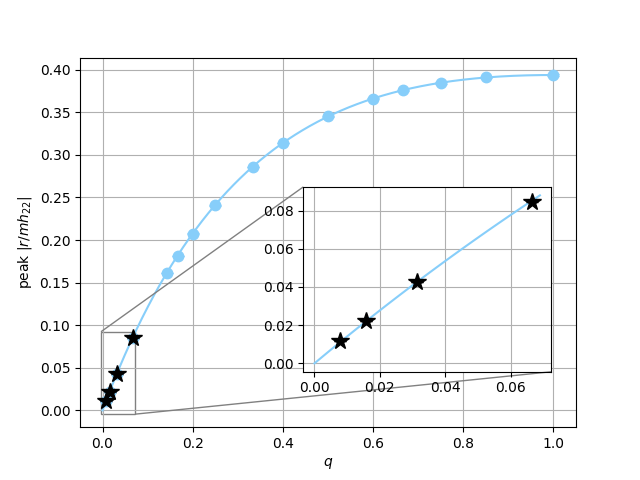}
\includegraphics[angle=0,width=0.98\columnwidth]{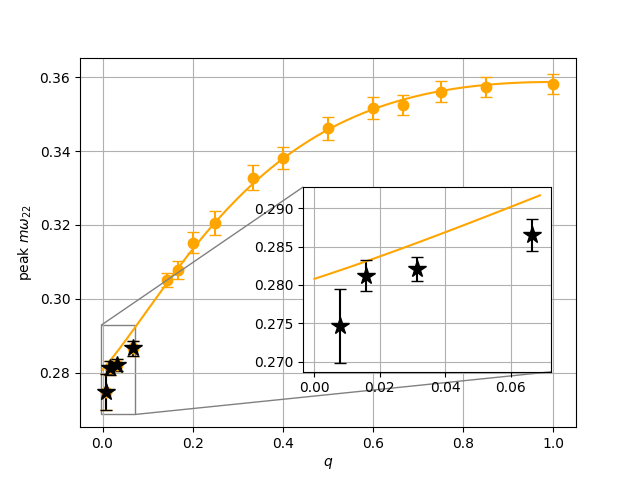}
\includegraphics[angle=0,width=0.98\columnwidth]{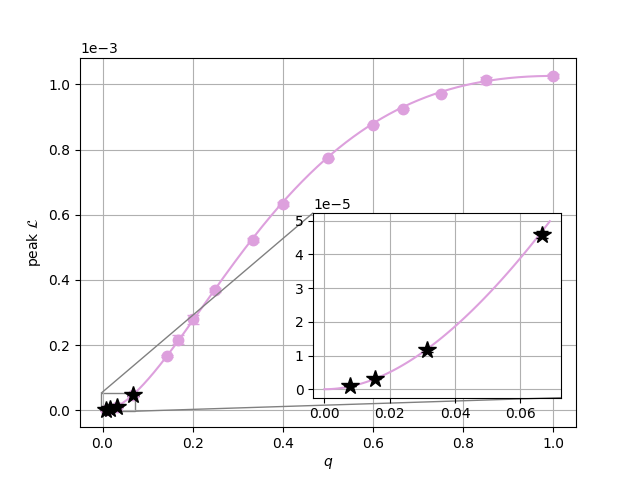}
  \caption{Final mass, spin, recoil velocity, 
peak amplitude, frequency, and luminosity. 
Predicted vs. current results for the 
$q=1/15, 1/32, 1/64, 1/128$ simulations. Each panel contains the prediction from the
   original fits in Ref.~\cite{Healy:2017mvh} (solid line), data used to determine
   the original fits (filled circles), and the data for the current results (stars).  An inset 
   in each panel zooms in on the new simulations.  Again, we stress no fitting to the new data
   is performed in this plot.
  \label{fig:predictions}}
\end{figure*}

An important test of accuracy of our simulations is to compare the
final properties in Table~\ref{tab:qqq} versus the predictions of the
formulas obtained in Ref.~\cite{Healy:2017mvh}
for them. We display these results in Fig.~\ref{fig:predictions}.
We stress that in this figure there is no fitting being performed to the new data,
but a raw comparison of the previous formulas extrapolated to a previously
uncovered region of small $q$. 

In the light of these good results,
we can now use the current data to generate a new fit of the
nonspinning binary remnant and merger waveform properties.
We will also correct for the center of mass motion\cite{Woodford:2019tlo,Healy:2020vre}
in these new fits,
of particular relevance for comparable masses, which was not 
included for all quantities in Ref.~\cite{Healy:2017mvh}.  In practice, the center of
mass motion for these nonspinning systems is small, and primarily only affects the
recoil velocity.  We find that after correction, the recoil velocities change by at most
2\%, well within the error from finite resolution (5-10\%).

For the unequal mass expansion of the final mass and spin of the merged
black holes we will use the forms from \cite{Healy:2014yta} that include the particle limit.
The fitting formula for $M_{\rm rem}$ is given by,
\begin{eqnarray}\label{eq:4mass}
\frac{M_{\rm rem}}{m} =& (4\eta)^2\,\Big\{M_0 +
                     K_{2d}\,\dmt^2 +
                     K_{4f}\,\dmt^4\Big\}\nonumber\\
                     &+\left[1+\eta(\tilde{E}_{\rm ISCO}+11)\right]\dmt^6,\quad\,
\end{eqnarray}
where $\delta{m}=(m_1-m_2)/m$ and $m=(m_1+m_2)$ and $4\eta=1-\delta{m}^2$.

And the fitting formula for the final spin has the form,
\begin{eqnarray}\label{eq:4spin}
\alpha_{\rm rem} = \frac{S_{\rm rem}}{M^2_{\rm rem}} =&
                     (4\eta)^2\Big\{L_0 +
                     L_{2d}\,\dmt^2+
                     L_{4f}\,\dmt^4\Big\}\nonumber\\
                     &+\eta\tilde{J}_{\rm ISCO}\dmt^6.
\end{eqnarray}
Note that the two formulae above impose the particle limit by including
the ISCO dependencies, $\tilde{E}_{\rm ISCO}(\alpha_{\rm rem})$, 
and $\tilde{J}_{\rm ISCO}(\alpha_{\rm rem})$ (See Ref. \cite{Healy:2014yta,Ori:2000zn} 
for the explicit expressions).

For the nonspinning recoil we will use Ref.~\cite{Healy:2017mvh} parametrization,
\begin{equation}\label{eq:vm}
v_m=\eta^2 \delta m\left(A+B\,\delta m^2+C\,\delta{m}^4\right).
\end{equation}

We model the peak
amplitude (of the strain $h$) from the merger of nonspinning
 binaries using the expansion \cite{Healy:2016lce},
\begin{eqnarray}\label{eq:4ph}
\left(\frac{r}{m}\right)h_{\rm peak} =& (4\eta)^2\,\Big\{H_0 +
                     H_{2d}\,\dmt^2 +
                     H_{4f}\,\dmt^4\Big\}\nonumber\\
                     &+\eta\,\tilde H_p\,\dmt^6,
\end{eqnarray}
where $\tilde H_p(\alpha_{\rm rem})$ is the particle limit, taking the value
$H_p(0)=1.4552857$ in the nonspinning limit~\cite{Bohe:2016gbl}.

The formula to model the peak luminosity introduced in \cite{Healy:2016lce}
takes the following simple form for nonspinning binaries,
\begin{equation}\label{eq:4plum}
{\cal L}_{\rm peak} = (4\eta)^2\,\Big\{N_0 +
                     N_{2d}\,\dmt^2 +
                     N_{4f}\,\dmt^4\Big\}.
\end{equation}

Analogously, we model the peak frequency
of the $(2,2)$ mode of the gravitational wave strain for 
nonspinning binaries as,
\begin{eqnarray}\label{eq:4omp}
m\omega_{22}^{\mathrm{peak}} =& (4\eta)\, \Big\{W_0 +
                     W_{2d}\,\dmt^2 +
                     W_{4f}\,\dmt^4\Big\}\nonumber\\
                     &+\tilde\Omega_p\,\dmt^6,
\end{eqnarray}
where $\tilde\Omega_p(\alpha_{\rm rem})$ is the particle limit, taking the value
$\tilde\Omega_p(0)=0.279525$ in the nonspinning limit~\cite{Bohe:2016gbl}.

\begin{table}
\caption{Fitting coefficients of the phenomenological formulas
(\ref{eq:4mass})-(\ref{eq:4omp})  \label{tab:coefficients}}
\begin{ruledtabular}
\begin{tabular}{lll}
$M_0$   &  $K_{2d}$ & $K_{4f}$\\
$0.95165\pm0.00002$ & $1.99604\pm0.00029$   & $2.97993\pm0.00066$\\
\hline
$L_0$   &  $L_{2d}$ & $L_{4f}$\\
$0.68692\pm0.00065$ & $0.79638\pm0.01086$   & $0.96823\pm0.02473$\\
\hline
$A$	& $B$	    &	$C$\\
$-8803.17\pm104.60$  &$-5045.58\pm816.10$   & $1752.17\pm1329.00$\\
\hline
$N_0\times10^{3}$    &  $N_{2d}\times10^{4}$    & $N_{4f}\times10^{4}$\\
$(1.0213\pm0.0004)$& $(-4.1368\pm0.0652)$      &$(2.46408\pm0.1485)$   \\
\hline
$W_0$  &  $W_{2d}$ & $W_{4f}$\\
$0.35737\pm0.00097$	& $0.26529\pm0.01096$ & $0.22752\pm0.01914$\\
\hline
$H_0$  &  $H_{2d}$ & $H_{4f}$\\
$0.39357\pm0.00015$& $0.34439\pm0.00256$  &  $0.33782\pm0.00584$\\
 \end{tabular}
 \end{ruledtabular}
 \end{table}

Table~\ref{tab:coefficients} summarizes the new values of all those coefficients with their estimated errors.

\section{Conclusions\label{sec:conclusions}}

This study represents a new Numerical Relativity milestone for 
comparative studies to assess
improvements in code efficiency, on gauge choices, and on improved initial
data, and it allows us to begin considering massive production of
small mass ratio simulations to populate the next release of
the RIT public catalog of binary black hole waveforms
(\url{https://ccrg.rit.edu/~RITCatalog}). In particular, the simulation
of mass ratio 128:1 presented here is a record breaking 
Numerical Relativity run
and it includes nearly 13 orbits before merger. The particle limit
remnant and peak waveform parameters are reproduced here within 1\%-2\% errors 
using purely full numerical methods and consistency of horizon and radiation
computations has been verified to high precision. 
The extension to include (high)
spins into the large black hole seems straightforward with our current
techniques~\cite{Ruchlin:2014zva} as well as to longer integrations times 
(with a fourth order Runge-Kutta method), where errors can be controlled by 
the reduction of the Courant factor. Those sort of improved simulations
could be used for direct parameter estimation of direct observations 
by advanced, 3rd generation 
ground based gravitational wave detectors and by the space project LISA.

\begin{acknowledgments}
The authors thank N.Rosato and Y.Zlochower for discussions,
and also gratefully acknowledge the National Science Foundation
(NSF) for financial support from Grant No.\ PHY-1912632. 
Local computational resources were provided by the NewHorizons, BlueSky
Clusters, and Green Prairies at the Rochester Institute of Technology,
which were supported by NSF grants No.\ PHY-0722703, No.\ DMS-0820923,
No.\ AST-1028087, No.\ PHY-1229173, and No.\ PHY-1726215.  
This work used the Extreme Science and Engineering Discovery 
Environment (XSEDE) [allocation TG-PHY060027N], which is supported by 
NSF grant No. ACI-1548562;
and the Frontera projects PHY-20010 and PHY-20007, an NSF-funded 
petascale computing system at the Texas Advanced Computing Center (TACC).
\end{acknowledgments}

\bibliography{../../../Bibtex/references}

\end{document}